\documentclass[authoryear,review,3p,10pt,onecolumn,leqno]{elsarticle}
\usepackage[english]{babel}

\usepackage{natbib}
\usepackage{url}
\usepackage{amsmath}
\usepackage{dsfont}
\usepackage{graphicx,graphics}
\usepackage{amssymb,amsmath}
\usepackage{verbatim,enumerate}
\usepackage{rotating}
\usepackage{multicol}
\usepackage{lmodern}
\usepackage{array}
\usepackage{multirow}
\usepackage{tabularx}
\usepackage{lscape, pdflscape}
\usepackage{url}
\usepackage{dcolumn}
\usepackage{setspace}
\RequirePackage{lineno}
\allowdisplaybreaks

\usepackage{epsfig, epstopdf}
\usepackage{color}
\usepackage{hyperref}
\usepackage{lineno}
\usepackage{amssymb,amsmath}
\usepackage{natbib}
\usepackage{setspace}
\usepackage{lscape}
\usepackage{afterpage,latexsym,epsfig,float}
\bibpunct{(}{)}{;}{a}{}{,}

\def\by{\mathbf{y}}
\def\bu{\mathbf{u}}
\def\bx{\mathbf{x}}
\def\bz{\mathbf{z}}
\def\bA{\mathbf{A}}
\def\bB{\mathbf{B}}
\def\bD{\mathbf{D}}
\def\bE{\mathbf{E}}
\def\bI{\mathbf{I}}
\def\bJ{\mathbf{J}}

\def\bT{\mathbf{T}}
\def\bX{\mathbf{X}}
\def\bY{\mathbf{Y}}
\def\bZ{\mathbf{Z}}

\newcommand{\bfeta}{\mbox{\protect\boldmath $\eta$}}

\def\bmu{\mathbf{\mu}}
\def\bnu{\mathbf{\nu}}

\def\bdelta{\mathbf{\delta}}

\def\bzero{\mathbf{0}}

\newtheorem{Rem}{Remark}

\newtheorem{proposition}{Proposition}
\newtheorem{Lem}{Lemma}
\newtheorem{Cor}{Corollary}
\newtheorem{Exa}{Example}

\makeatletter
\renewcommand\@biblabel[1]{#1.}
\makeatother

\journal{Physica A}

\begin{document}

\bibliographystyle{siam}

\title{R\'enyi entropy and complexity measure for\\
skew-gaussian distributions and related families}

\author[IFOP,UTFS]{Javier E. Contreras-Reyes\corref{cor1}}
\ead{jecontrr@uc.cl, javier.contreras@ifop.cl}

\cortext[cor1]{Corresponding author, Phone +56 032 2151 513}
\address[IFOP]{Division of Fisheries Research, Fisheries Development Institute, Blanco 839, Valpara\'iso, Chile}
\address[UTFS]{Department of Mathematics, Universidad T\'ecnica Federico Santa Mar\'ia, Valpara\'iso, Chile}

\begin{abstract}
In this paper, we provide the R\'enyi entropy and complexity measure for a novel, flexible class of skew-gaussian
distributions and their related families, as a characteristic form of the skew-gaussian Shannon entropy. We give
closed expressions considering a more general class of closed skew-gaussian distributions and the weighted moments
estimation method. In addition, closed expressions of R\'enyi entropy are presented for extended skew-gaussian and
truncated skew-gaussian distributions. Finally, additional inequalities for skew-gaussian and extended skew-gaussian
R\'enyi and Shannon entropies are reported.\\

\end{abstract}

\maketitle

\section{Introduction}

The family of skew-gaussian distributions has been popularized by \citet{Azzalini_1985}
and ever since it has been discussed extensively in the literature. Such discussions include a wide variety of
skewed models in addition to having gaussian distribution as a special case and flexibility in capturing
skewness in the data \citep{Azzalini_Dalla-Valle_1996,Azzalini_Capitanio_1999,Azzalini_2013}. In this
sense, \citet{Gonzalez-Farias_et_al_2004} present the closed skew-gaussian distribution as an extension
of the skew-gaussian case, but closed under operations such as sums, marginalization, and linear
conditioning \citep{Rezaie_et_al_2014}. Another generalization of the skew-gaussian distribution is the
extended skew-gaussian distribution \citep{Capitanio_et_al_2003} that adds a fourth real parameter to
accommodate both skewness and heavy tails. In some cases where observed variables can be simultaneously
skewed and restricted to a fixed interval, the truncated skew-gaussian distribution is a good choice for
those applications, especially for environmental and biological variables in which the observations are
positives \citep{Flecher_et_al_2010}.

In many applications, the empirical distribution of some observed variables was modeled by a skew-gaussian
distribution. For example, the closed skew-gaussian distribution is used by \citet{Rezaie_et_al_2014} to
simulate seismic amplitude variations. \citet{Contreras-Reyes_Arellano-Valle_2012} consider the skew-gaussian
distribution for seismic magnitudes of aftershocks catalogue of the 2010 Maule earthquake in Chile;
\citet{Arellano-Valle_et_al_2013} for the optimization of ozone's monitoring network; and \citet{Figiel_2014}
for a digital reconstruction of nanocomposite morphologies from TEM (Transmission Electron Microscopy) images.
An implementation of the extended skew-gaussian (in logarithmic form) can be found in \citet{Zhou_Wang_2008}
for pricing of both Asian and basket options. As mentioned above, \citet{Flecher_et_al_2010} considers the
truncated skew-gaussian distribution to fit the daily relative humidity measurements. See more applications
in \citet{Genton_2004}.

More recently, \citet{Contreras-Reyes_Arellano-Valle_2012} and \citet{Arellano-Valle_et_al_2013} compute the
Kullback-Leibler divergence measure for skew-gaussian distribution and Shannon entropy for the full class of
skew-elliptical distributions, respectively. They highlight that the Kullback-Leibler information measure should be
represented in quadratic form, including a non-analytical expected value. In addition, they gave the Kullback-Leibler
divergence of a multivariate skew-gaussian distribution with respect to multivariate gaussian distribution. Information
measure applications dealing with skewed data have been performed by \citet{Contreras-Reyes_Arellano-Valle_2012},
\citet{Arellano-Valle_et_al_2013}, \citet{Contreras-Reyes_2014} and references therein.

In this work, we focus on the R\'enyi entropy \citep{Renyi_1970} as a characteristic form of the Shannon entropy to
give a closed expression of skew-gaussian densities. Additionally, the LMC complexity measure \citep{Lopez-Ruiz_et_al_1995}
is derived by the difference between the extensive R\'enyi entropy and Shannon entropy \citep{Yamano_2004}. To do this,
we briefly describe the main properties of closed skew-gaussian distributions. Finally, we compute the R\'enyi entropy
and complexity measure for the extended skew-gaussian and univariate truncated skew-gaussian densities.\\
%Some proofs are presented in Appendix.

\section{R\'enyi entropy and complexity measure}

Consider the $\alpha$th-order R\'enyi entropy \citep{Renyi_1970} of
probability density $f(\bx)$ on a variable $\bx\in\Delta\subset\mathds{R}^d$:
\begin{equation}\label{red}
R_{\alpha}[f]=\frac{1}{1-\alpha}\ln\int [f(\bx)]^{\alpha}d\bx,
\end{equation}
where normalization to unity as given by $\int f(\bx)d\bx=1$ \citep{Sanchez-Moreno_et_al_2014}.
\citet{Golshani_Pasha_2010} provide some important properties of the R\'enyi entropy: 1. $R_{\alpha}[f]$
can be negative, 2. $R_{\alpha}[f]$ is invariant under a location transformation, 3. $R_{\alpha}[f]$
is not invariant under a scale transformation, and 4. for any $\alpha_1<\alpha_2$, $\bx\in\Delta$,
we have $R_{\alpha_1}[f]\geq R_{\alpha_2}[f]$, which are equal if and only if $\bx$ is uniformly
distributed.

From (\ref{red}), the Shannon entropy is obtained by the limit
\begin{equation}\label{sha}
S[f]=\lim_{\alpha\rightarrow 1}R_{\alpha}[f]=-\int f(\bx)\ln f(\bx)d\bx
\end{equation}
by applying l'H\^opital's rule to $R_{\alpha}[f]$ with respect to $\alpha$ \citep{Renyi_1970}.
This measure is the expected value of $g(\bx)=-\ln f(\bx)$ with respect to $f(\bx)$, i.e.,
$S[f]=\langle g(\bx)\rangle$ \citep{Liu_et_al_2012}. Hereafter, we will refer to this as the expected
information of $g(\bx)$ in $\bx$. See \citet{Cover_Thomas_2006} for additional properties of
the Shannon entropy.

\begin{Exa} \citep{Dembo_et_al_1991,Cover_Thomas_2006}.
Let $\bx$ be gaussian with mean vector $\bmu\in\mathds{R}^d$ and $\bJ$ is a $d\times d$ variance matrix
(with determinant $|\bJ|>0$). Then, the R\'enyi and Shannon entropies of $\bx$ are given by
\begin{eqnarray}
R_{\alpha}[f]&=&\frac{1}{2}\ln[(2\pi)^d|\bJ|] + \frac{d\ln\alpha}{2(\alpha-1)},\quad 1<\alpha<\infty,\label{rn}\\
S[f]&=&\frac{1}{2}\ln[(2\pi e)^d|\bJ|],\label{sn}
\end{eqnarray}
respectively.
\end{Exa}

Another important concept is the statistical complexity that measures the randomness and
structural correlations of a known system \citep{Carpi_et_al_2011}. \citet{Lopez-Ruiz_et_al_1995}
proposed a measure of statistical complexity (LMC) in order to determine the {\it disequilibrium}
of the system attributed to entropy measure \citep{Anteneodo_Plastino_1996,Sanchez-Moreno_et_al_2014}.
LMC measure is defined as the product
\begin{equation}\label{LMC}
C_{LMC}[f]=e^{S[f]-R_2[f]},
\end{equation}
where $R_2[f]$ is the quadratic R\'enyi entropy of $\bx$ ($\alpha=2$). \citet{Yamano_2004} provide
an extensive entropy instead of an additive Shannon entropy in (\ref{LMC}), characterised as a
difference between the $\alpha$th-order R\'enyi entropy and quadratic R\'enyi entropy as
\begin{equation}\label{CM}
C_{\alpha}[f]=e^{R_{\alpha}[f]-R_2[f]}.
\end{equation}
Note that $C_{\alpha}[f]$ reflects the shape of the distribution of $\bx$ and takes unity for all
distributions when $\alpha=2$. In addition, $\mathcal{C}_{\alpha}$ satisfies a great variety of interesting
mathematical and physical properties. Let us just recall here the following properties: 1. $C_{\alpha}[f]>1$,
$\forall\,\alpha\leq2$, and, $0<C_{\alpha}[f]\leq1$, $\forall\,\alpha>2$; 2. $C_{\alpha}[f]$ is invariant under
a location and scale transformation in the distribution of $\bx$; and 3. is invariant under replications
of the original distribution of $\bx$.\\

\section{Skew-gaussian distribution and related families}

The closed skew-gaussian distribution has interesting properties inherited from the Gaussian distribution
and corresponds to a generalization of the skew-gaussian distribution. We briefly describe some of its inferential
properties and present the weighted moments method in Proposition~2 \citep{Flecher_et_al_2009}, necessary to
calculate R\'enyi entropy of skew-gaussian random vectors.\\

\subsection{Closed skew-gaussian distributions}

Concerning the definition of \citet{Flecher_et_al_2009} and \citet{Gonzalez-Farias_et_al_2004}, let
$\by\in\Delta\subset\mathds{R}^d$ be a random vector with closed skew-gaussian distribution denoted as
$CSN_{d,s}(\bmu,\bJ,\bD,\nu,\bA)$ and with density function
\begin{equation}\label{csn}
f_{d,s}(\by)=\phi_d(\by;\bmu,\bJ)\frac{\Phi_s(\bD^{\top}(\by-\bmu);\bnu,\bA)}{\Phi_s(\bzero;\bnu,\bA+\bD^{\top}\bJ\bD)},
\end{equation}
where $\bmu\in\mathds{R}^d$, $\bnu\in\mathds{R}^s$, $\bJ\in\mathds{R}^{d\times d}$ and $\bA\in\mathds{R}^{s\times s}$
are both covariance matrices, $\bD\in\mathds{R}^{d\times s}$, $\bD^{\top}$ denotes the transposed $\bD$ matrix,
$$\phi_d(\by;\bmu,\bJ)=\frac{1}{(2\pi)^{d/2}|\bJ|^{1/2}}\exp\left(-\frac{1}{2}(\by-\bmu)^{\top}\bJ^{-1}(\by-\bmu)\right)$$
and
$\Phi_d(\by;\bmu,\bJ)$ are the probability function (pdf) and cumulative distribution function, respectively, of the
$d$-dimensional gaussian distribution with mean vector $\bmu$ and variance matrix $\bJ$. The closed skew-gaussian
distribution is closed under translations, scalar multiplications, and full, row rank linear transformations
\citep{Gonzalez-Farias_et_al_2004,Genton_2004}. Let $\bT\in\mathds{R}^{n\times d}$ be a matrix with rank $n$ such that
$d\leq n$, then
\begin{equation}\label{acsn}
\bT\by=CSN_{n,s}(\bT\bmu,\tilde{\bJ},\tilde{\bD},\bnu,\tilde{\bT})
\end{equation}
where $\tilde{\bJ}=\bT^{\top}\bJ\bT$, $\tilde{\bD}=\bD^{\top}\bJ \bT\tilde{\bJ}^{-1}$, and
$\tilde{\bT}=\bA + \bD^{\top}\bJ\bD-\tilde{\bD}^{\top}\tilde{\bJ}\tilde{\bD}$
\citep[][see Proposition 2.3.1]{Genton_2004}.

A particular case of (\ref{acsn}), is the standardised random vector $\bz_0=\bJ^{-1}(\by-\bmu)$. In this case,
Eq. (\ref{csn}) is rewritten as
\begin{equation}\label{scsn}
f_{d,s}(\bz_0)=\phi_d(\bz_0)\frac{\Phi_s(\bD^{\top}\bJ^{1/2}\bz_0;\bnu,\bA)}{\Phi_s(\bzero;\bnu,\bA + \bD^{\top}\bJ\bD)}.
\end{equation}

Given that the closed skew-gaussian distribution is closed under translations and by property (\ref{acsn}), the
standardised random vector $\bZ_0$ follows $CSN_{d,s}(\bzero,\bI_d,\bD^{\top}\bJ^{1/2},\bnu,\bA)$,
where $\bI_d$ denotes the $d$-dimensional identity matrix. For more details, see \citet{Flecher_et_al_2009} and
\citet{Genton_2004}.
%For the moment generating function of the CSN distribution, see \cite{Gonzalez-Farias_et_al_2004}.

\begin{Lem}\label{Pf} \citep{Flecher_et_al_2009}.
Let $\bY$ be a $CSN_{d,s}(\bmu,\bJ,\bD,\bzero,\bA)$, $r$ a positive integer and
$h(\by)=h(y_1,\ldots,y_d)$ be any real valued function such that $\langle h(\bY)\rangle$
is finite, then
\begin{equation}\label{ew}
\langle h(\bY)[\Phi_d(\bY;\bzero,\bI_d)]^r\rangle=\langle h(\tilde{\bY})\rangle
\frac{\Phi_{rd+s}(\bzero;\tilde{\bnu},\tilde{\bA}+\tilde{\bD}^{\top}\bJ\tilde{\bD})}{\Phi_s(\bzero;\bzero,\bA+\bD^{\top}\bJ\bD)},
\end{equation}
where $\tilde{\bY}\sim CSN_{d,rd+s}(\bmu,\bJ,\tilde{\bD},\tilde{\bnu},\tilde{\bA})$ with
$\tilde{\bD}=(\bE^{\top},\,\bD^{\top})$, $\bE$ a $d\times rd$ matrix defined by $\bE=(\bI_d,\ldots,\bI_d)$,
$\tilde{\bnu}=(-\bmu,\ldots,-\bmu,\bzero_s)$ a $(rd+s)$ vector and $$\tilde{\bA}=\left(
                     \begin{array}{cc}
                       \bI_{rd} & \bzero  \\
                       \bzero & \bA \\
                     \end{array}
                   \right).$$
\end{Lem}

\subsection{Skew-gaussian distribution}

A special case of closed skew-gaussian is the gaussian density when $\bD=\bzero$. When $s=1$, the skew-gaussian density function
is obtained \citep{Azzalini_Dalla-Valle_1996,Azzalini_Capitanio_1999,Azzalini_2013}. For simplicity, a slight variant of the original
definition is considered here. In this work it is posited that a random vector $\bZ\in\Delta\subset\mathbb{R}^d$ has a skew-gaussian
distribution with mean vector $\bmu\in\mathbb{R}^d$, variance matrix $\bJ\in\mathbb{R}^{d\times d}$ and shape/skewness parameter
$\bfeta\in\mathbb{R}^d$, denoted by $\bZ\sim SN_d(\bmu,\bJ,\bfeta)$, if its probability density function is
\begin{eqnarray}
f(\bz)=2\phi_d(\bz;\bmu,\bJ)\Phi_1[\bfeta^{\top}(\bz-\bmu)].\label{SN-pdf}
\end{eqnarray}

The mean vector and the variance matrix of $\bZ$ are
\begin{eqnarray}
\langle \bz\rangle&=&\bmu+\sqrt{\frac{2}{\pi}}\,\bdelta,\nonumber\\
\langle \bz^2\rangle&=&\bJ-\frac{2}{\pi}\bdelta\bdelta^{\top},\nonumber
\end{eqnarray}
respectively, where $\bdelta=\bJ\bfeta/\sqrt{1+\bfeta^{\top}\bJ\bfeta}$
\citep{Azzalini_Capitanio_1999,Contreras-Reyes_Arellano-Valle_2012}.

\begin{proposition}\label{T0}
Let $\bZ$ be a $SN_d(\bmu,\bJ,\bfeta)$. Then:
\begin{equation}\label{re1}
\int [f(\bz)]^{\alpha}d\bz=\psi_{\alpha,d}(\bJ)\,
\frac{\Phi_{\alpha+1}(\bzero;\bzero,\tilde{\bJ})}{\Phi_1(0;0,\sigma^2)},\quad\alpha\in\mathbb{N},\,\alpha>1,
\end{equation}
where
$$\psi_{\alpha,d}(\bJ)=\frac{2^{\alpha}}{\alpha^{d/2}}[(2\pi)^d|\bJ|]^{(1-\alpha)/2},$$
$\tilde{\bJ}=\bI_{\alpha+1}+\|\tilde{\bfeta}\|^2\tilde{\bD}^{\top}\tilde{\bD}$,
$\tilde{\bD}=({\bf 1}_{\alpha},\|\tilde{\bfeta}\|)^{\top}$, ${\bf 1}_{\alpha}$ is the $\alpha$-dimensional
vector of ones, $\sigma^2=1+\|\tilde{\bfeta}\|^4$, $\|\tilde{\bfeta}\|=\tilde{\bfeta}^{\top}\tilde{\bfeta}$
and $\tilde{\bfeta}=\alpha^{-1/2}\bJ^{1/2}\bfeta$.
\end{proposition}

By (\ref{red}) and (\ref{re1}), the R\'enyi entropy of a random variable $\bZ\sim SN_d(\bmu,\bJ,\bfeta)$ is
retrieved. Taking $\bfeta=\bzero$ in (\ref{re1}), the R\'enyi entropy of the gaussian distribution given by
(\ref{rn}) is obtained. Lemma~\ref{Pf} allows the computing of the expected value of the cumulative density
function of a gaussian density. Considering the standarised closed skew-gaussian variable in (\ref{scsn}),
the Proposition~\ref{T0} is solved by (\ref{ew}), by setting $\bnu=\bzero$ and $\bA=\bI_d$, with $d=s=1$.
However, the case $\bnu\neq\bzero$ and $\bA\neq\bI_d$, $d>1$, is still an open problem and, it is useful to
find the R\'enyi entropy for closed skew-gaussian distributions. By (\ref{red}) and (\ref{scsn}), the
Shannon entropy for closed skew-gaussian distributions is rewritten as
\begin{eqnarray}
S[f]&=&-\langle\ln[f_{d,s}(\bY)]\rangle\nonumber\\
&=&\frac{1}{2}\ln|\bJ|-\ln[\Phi_s(\bzero;\bnu,\bA + \bD^{\top}\bJ\bD)] - \langle\ln[\phi_d(\bZ_0)\Phi_s(\tilde{\bD}^{\top}\bZ_0;\bnu,\bA)]\rangle\nonumber\\
&=&S[f_0]-\ln[\Phi_s(\bzero;\bnu,\bA + \bD^{\top}\bJ\bD)] - \langle\ln[\Phi_s(\tilde{\bD}^{\top}\bZ_0;\bnu,\bA)]\rangle\label{ShCSN},
\end{eqnarray}
where $f_0$ is the standardised gaussian distribution and $S[f_0]=(1/2)\ln(2\pi e)$.

\begin{Cor}\label{NRH}
Let $\bZ\sim SN_d(\bmu,\bJ,\bfeta)$, $\bZ_N\sim N_d(\bmu,\bJ)$, $\|\tilde{\bfeta}\|=\tilde{\bfeta}^{\top}\tilde{\bfeta}$ and $\tilde{\bfeta}=\bJ^{1/2}\bfeta$. Then,
\begin{itemize}
\item[(i)] $\displaystyle\begin{aligned}[t]
R_{\alpha}[f]&=R_{\alpha}[f_0]-N_{\alpha}[f],\quad\alpha\in\mathbb{N},\,\alpha>1,\,\mbox{where}\\
\end{aligned}$

 $$N_{\alpha}[f]=\frac{1}{\alpha-1}{\rm ln}\left[2^{\alpha}\frac{\Phi_{\alpha+1}(\bzero;\bzero,\tilde{\bJ})}{\Phi_1(0;0,\sigma^2)}\right]$$
is the so-called {\it Negentropy}, $R_{\alpha}[f_0]$ is given by (\ref{rn}), and $\tilde{\bJ}$ and $\sigma^2$ are defined
as in Proposition~\ref{T0}.

\item[(ii)] $\displaystyle\begin{aligned}[t]
\lim_{\alpha\rightarrow 1}N_{\alpha}[f]=\langle{\rm ln}[2\Phi_1(\|\tilde{\bfeta}\|W)]\rangle.
\end{aligned}$

\item[(iii)] $\displaystyle\begin{aligned}[t]
S[f]&=S[f_0]-\langle{\rm ln}[2\Phi_1(\|\tilde{\bfeta}\|W)]\rangle,
\end{aligned}$

where $S[f_0]$ is given by (\ref{sn}) and $W\sim SN_1(0,1,\|\tilde{\bfeta}\|)$.

\item[(iv)] $\displaystyle\begin{aligned}[t]
S[f_0] - {\rm ln}(4e)&\leq S[f] \leq S[f_0],\,\forall\,\bfeta.\\
\end{aligned}$
\end{itemize}
\end{Cor}

\citet{Contreras-Reyes_Arellano-Valle_2012} define the negentropy as the departure from gaussianity of the
distribution of $\bZ$. Therefore, the skew-gaussian R\'enyi entropy corresponds to the difference between
gaussian R\'enyi entropy and negentropy, that depends on the skewness parameter $\bfeta$. On the another
hand, by setting $\bnu=\bzero$ and $\bA=\bI_d$ in (\ref{ShCSN}) with $d=s=1$, we obtain the property
(ii) of Corollary~\ref{NRH}.

By properties (iii) and (iv),
$$-0.967\leq S[f_0]-{\rm log}\,(4e)\leq S[f]$$
because, the minimum value of normal Shannon entropy is obtained for $d=1$ and,
$$0\leq\langle\ln[2\Phi_1(\|\tilde{\bfeta}\|W)]\rangle\leq 2.386,$$
for all $\bfeta$. In addition, \citet{Contreras-Reyes_Arellano-Valle_2012} reported a maximum value of this
expected value equal to 2.339, using numerical approximations. Considering (\ref{red}), (\ref{CM}) and
(\ref{re1}); the complexity measure for skew-gaussian distribution is obtained.\\

\subsection{Extended skew-gaussian distributions}

Consider a slight variant of the extended skew-gaussian distribution proposed by \citet{Capitanio_et_al_2003}. Let
$\bZ\sim ESN_d(\bmu,\bJ,\bfeta,\tau)$, $\bZ\in\Delta\subset\mathbb{R}^d$, with mean vector $\bmu\in\mathbb{R}^d$,
variance matrix $\bJ\in\mathbb{R}^{d\times d}$, shape/skewness parameter $\bfeta\in\mathbb{R}^d$, extended
parameter $\tau\in\mathbb{R}$, and with pdf given by:
\begin{equation}\label{dmesn}
p(\bz)=\frac{1}{\Phi_1(\tau)}\phi_d(\bz;\bmu,\bJ)\Phi_1[\bfeta^{\top}(\bz-\bmu)+\tilde{\tau}],
\end{equation}
where $\bz\in\mathbb{R}^d$ and $\tilde{\tau}=\tau\,\sqrt{1+\bfeta^{\top}\bJ\bfeta}$.
The mean vector and the variance matrix of $\bZ$ are
\begin{eqnarray}
\langle\bz\rangle&=&\bmu+\bdelta\zeta_1(\tau),\label{esn-mom1}\\
\langle\bz^2\rangle&=&\bJ-\zeta_1(\tau)[\tau+\zeta_1(\tau)]\bdelta\bdelta^\top,\label{esn-mom2}
\end{eqnarray}
respectively; where $\zeta_1(\bz)=\phi(\bz)/\Phi_1(\bz)$ is the {\it zeta} function
\citep{Azzalini_Capitanio_1999,Capitanio_et_al_2003}.

%From \cite{Arellano-Valle_Azzalini_2006}, a stochastic representation of the
%$ESN_d$ distribution is
%\begin{equation}\label{sn-sr}
%\bZ\buildrel d\over=\bW + \bdelta U,
%\end{equation}
%where $\bdelta=\bOmega\bfeta/\sqrt{1+\bfeta^{\top}\bOmega\bfeta}$, $U\sim LTN_{(-\tau,\infty)}(0,1)$,
%which is independent of $\bW\sim N_d\left(\bxi,\bJ\right)$, $\bJ=\bOmega-\bdelta\bdelta^{\top}$,
%where $LTN_{(-\tau,\infty)}(0,1)$ represents the unit normal distribution truncated below the point $-\tau$
%and ``$\buildrel d\over=$'' denotes equality in terms of distribution. From the stochastic representation
%(\ref{sn-sr}) it follows that $\bZ\buildrel d\over=\bW+\bdelta W_{\tau}$, where
%$W_{\tau}\buildrel d\over=(W_{0}\mid W_{0}+\tau>0)$ and
%        \begin{equation*}
%        \left(%
%        \begin{array}{c}
%        W_{0} \\
%        \bW\\
%        \end{array}%
%        \right)
%        \sim N_{1+d}\left(
%        \left(%
%        \begin{array}{c}
%        0 \\
%        \bxi \\
%        \end{array}%
%        \right),\left(%
%        \begin{array}{cc}
%        1 & {\bf 0}^\top \\
%        {\bf 0} & \bJ \\
%        \end{array}%
%        \right)\right).
%        \end{equation*}
%Note that $W_{0}$ and $\bW$ are independent. This fact must be interpreted as
%the non-normality effect of $W_{0}$ on $\bZ$, where $W_{0}$ is the so-called
%{\it unobserved confounder} variable. From (\ref{sn-sr}),

\begin{proposition}\label{T2}
Let $\bZ$ be a $ESN_d(\bmu,\bJ,\bfeta,\tau)$, $\bz\in\mathbb{R}^d$. Then:
\begin{equation}\label{re3}
\int [f(\bz)]^{\alpha}d\bz=\psi_{\alpha,d}(\bJ)\langle\left[\frac{\Phi_1(W)}{2\Phi_1(\tau)}\right]^{\alpha}\rangle,\quad\alpha\in\mathbb{N},\,\alpha>1,
\end{equation}
where $\psi_{\alpha,d}(\bJ)$ is defined as in Proposition~\ref{T0} and $W={\tilde{\bfeta}}^{\top}\bZ_0+\tilde{\tau}\sim ESN_1(\tilde{\tau},\|\tilde{\bfeta}\|^2,\|\tilde{\bfeta}\|,\tau)$, $\|\tilde{\bfeta}\|=\tilde{\bfeta}^{\top}\tilde{\bfeta}$,
and $\tilde{\bfeta}=\bJ^{1/2}\bfeta$.
\end{proposition}

\begin{Cor}\label{NHESN}
Let $\bZ\sim ESN_d(\bmu,\bJ,\bfeta,\tau)$, $\bZ_N\sim N_d(\bmu,\bJ)$ and
$W$ are defined as in Proposition~\ref{T2}. Then,
\begin{itemize}
\item[(i)] $\displaystyle\begin{aligned}[t]
R_{\alpha}[f]&=R_{\alpha}[f_0]-N_{\alpha}[f],\quad\alpha\in\mathbb{N},\,\alpha>1,\\
\end{aligned}$

where $$N_{\alpha}[f]=\frac{1}{\alpha-1}{\rm ln}\langle\left[\frac{\Phi_1(W)}{\Phi_1(\tau)}\right]^{\alpha}\rangle,$$
and $R_{\alpha}[f_0]$ is given by (\ref{rn}).

\item[(ii)] $\displaystyle\begin{aligned}[t]
R_{\alpha}[f]&\leq R_{\alpha}[f_0] + \frac{\alpha}{1-\alpha}{\rm ln}\left[\frac{\Phi_1(\tilde{\tau} + \tilde{\delta}\zeta_1(\tau))}{\Phi_1(\tau)}\right],\\
\end{aligned}$

where $\tilde{\delta}=\|\tilde{\bfeta}\|^3/\sqrt{1+\|\tilde{\bfeta}\|^4}$.

\item[(iii)] $\displaystyle\begin{aligned}[t]
S[f]&=S[f_0]-\langle{\rm ln}\left[\frac{\Phi_1(W)}{\Phi_1(\tau)}\right]\rangle.\\
\end{aligned}$

\item[(iv)] $\displaystyle\begin{aligned}[t]
S[f_0]+{\rm ln}[\Phi_1(\tau)]-\Phi_1\left(\frac{\tilde{\tau}}{\sqrt{1+\bfeta\bfeta^\top}}\right)&\leq
S[f]\leq \frac{1}{2}{\rm ln}\left[(2\pi e)^d\left|\bJ-\zeta_1(\tau)[\tau+\zeta_1(\tau)]\bdelta\bdelta^\top\right|\right],\, \forall\,\bfeta.\\
\end{aligned}$

\item[(v)] $\displaystyle\begin{aligned}[t]
\lim_{\alpha\rightarrow 1}N_{\alpha}[f]=\langle{\rm ln}[\frac{\Phi_1(W)}{\Phi_1(\tau)}]\rangle.
\end{aligned}$
\end{itemize}
\end{Cor}

\citet{Pourahmadi_2007} illustrated the behaviour of $\zeta_1(\tau)$, $\tau\in\mathbb{R}$. This function is strictly decreasing
for any $\tau\in\mathbb{R}$, tends to 0 when $\tau\rightarrow +\infty$, and diverge when $\tau\rightarrow -\infty$. For $\tau=0$,
the property (iv) of Corollary~\ref{NHESN} becomes property (iii) of Corollary~\ref{NRH}. By properties (iii) of Corollary~\ref{NHESN}
and (ii) of Corollary~\ref{NRH}, the negentropy of an extended skew-gaussian random vector is always larger than the negentropy of a
skew-gaussian random vector. Therefore, we obtain the following relationship among the Shannon entropies of gaussian ($f_0(\bz)$),
skew-gaussian ($g(\by)$), and extended skew-gaussian ($f(\bx)$) distributions: $S[f_0]\geq S[g]\geq S[f]$. Considering (\ref{red}),
(\ref{CM}) and (\ref{re3}); the complexity measure for extended skew-gaussian distribution is obtained.\\

\subsection{Truncated skew-gaussian distributions}

The truncated skew-gaussian pdf given by \citet{Flecher_et_al_2010}, consider the random variable $Z\sim SN_1(\mu,\omega,\lambda)$,
$\bZ\in\Delta\subset\mathbb{R}$, and the definition given in (\ref{SN-pdf}) for the case $d=1$. \citet{Flecher_et_al_2010} gives
the expressions of the higher order and weighted moments of truncated skew-gaussian distributions. We also consider the following
definition based on (\ref{SN-pdf}) for a truncated skew-gaussian random variable $W\in[a,b]\subset\mathbb{R}$, denoted by
$W\sim TSN(\mu,J,\lambda)$, and with density
\begin{equation}\label{TSN}
g(w)=\frac{f(w)}{[F(w)]_a^b}, \quad\mbox{$a<w\leq b$},
\end{equation}
where $f(z)$ is defined in (\ref{SN-pdf}) for $d=1$ with $\bJ=J$, $\bfeta=\lambda$; and $F(z)$ is the cumulative density function
of $Z$ with
$$[F(w)]_a^b=F(b)-F(a)=\int_{a}^{b}f(u)du.$$

The following Remark allows the computation of $[F(w)]_a^b$ in terms of the gaussian cumulative density
function and a bivariate integral term.

\begin{Rem}
Let $Z\sim SN_1(\mu,J,\lambda)$, \citet{Owen_1956} and \citet{Azzalini_1985} gives the
expressions to compute $F(z)$ as follows
\begin{equation}\label{SNcdf}
F(z)=2\int_{z}^{-\infty}\int_{-\infty}^{\lambda s}\phi(s)\phi(t)\,dt\,ds=\Phi_1(z)-2\int_{z}^{\infty}\int_{0}^{\lambda s}\phi(s)\phi(t)\,dt\,ds.
\end{equation}
Then, by replacing (\ref{SNcdf}) in $[F(w)]_a^b$ we obtain
$$[F(w)]_a^b=\Phi_1(b)-\Phi_1(a)-2\int_{a}^{b}\int_{0}^{\lambda s}\phi(s)\phi(t)dtds.$$
\end{Rem}

%Note in (\ref{SNcdf2}) that the behaviour of the left and right tails of a skew-normal distribution
%related to $X\sim SN_1(\mu,\omega,\lambda)$ is $T(a;\lambda)=T(b;\lambda)=0$, when $a\to-\infty$ and
%$b\to+\infty$, for any $\lambda$ and; $T(a;0)=T(b;0)=0$ for any truncations $a$ and
%$b$ parameters in the left and right tails, respectively \cite{Owen_1956}.

\begin{proposition}\label{T1}
Let $Z,\,W$ be a $SN_1(\mu,J,\lambda)$ and $TSN_1(\mu,J,\lambda)$, respectively, $\lambda\neq0$. Then:
\begin{equation}\label{re2}
\int_a^b [g(w)]^{\alpha}dw=2\psi_{\alpha,1}(J)\,\Phi_{\alpha+1}(\bzero;\bzero,\tilde{\bJ})\frac{[H(v)]_{a_0}^{b_0}}{([F(z)]_{a}^{b})^{\alpha}},
\end{equation}
where $\psi_{\alpha,1}(J)$ is defined as in Proposition~\ref{T0} with $d=1$ and $\bJ=J$;
$\tilde{\bJ}=\bI_{\alpha+1}+\tilde{\lambda}^2\tilde{\bD}^{\top}\tilde{\bD}$, $\tilde{\lambda}^2=\omega\lambda^2/\alpha$,
$\tilde{\bD}=({\bf 1}_{\alpha},\tilde{\lambda})^{\top}$ and $V\sim CSN_{1,2}(0,\tilde{\lambda}^2,\tilde{\bB},\bzero,\bI_2)$
with cumulative density function $H(v)$, $\tilde{\bB}=(1,\tilde{\lambda})^{\top}$, $a_0=\lambda(a-\mu)/\omega$ and $b_0=\lambda(b-\mu)/\omega$.
\end{proposition}

\begin{Rem}
By Lemma~2.2.1 of \citet{Genton_2004}, $H(v)$ is easily computable by a tri-variate gaussian cumulative density function as
\begin{eqnarray*}
H(v)&=&\frac{\Phi_3\left[\left(
                     \begin{array}{c}
                       v \\
                       {\bf 0} \\
                     \end{array}
                   \right);
                   \left(
                     \begin{array}{c}
                       0 \\
                       \bzero \\
                     \end{array}
                   \right),
                   \left(
                     \begin{array}{ccc}
                       \tilde{\lambda}^2 & - \tilde{\lambda}^2\tilde{\bB}  \\
                       - \tilde{\lambda}^2\tilde{\bB}^{\top} & \bI_2 + \tilde{\lambda}^2\tilde{\bB}^{\top}\tilde{\bB} \\
                     \end{array}
                   \right)\right]}{\Phi_2({\bf 0};{\bf 0},\bI_2 + \tilde{\lambda}^2\tilde{\bB}^{\top}\tilde{\bB})},
\end{eqnarray*}
where $\tilde{\lambda}$ and $\tilde{\bB}$ are defined as in Proposition~\ref{T1}.
\end{Rem}

Considering (\ref{red}), (\ref{CM}) and (\ref{re2}); the complexity measure for
extended skew-gaussian distribution is obtained.\\

\section{Conclusions}

In this paper, we have presented some solutions to compute the R\'enyi entropy with discrete $\alpha$-order
and for a wide range of asymmetric distributions. Specifically, we find a closed expression
for skew-gaussian, extended skew-gaussian, and truncated skew-gaussian distributions. Finally, additional
inequalities for skew-gaussian and extended skew-gaussian entropies were reported.\\

%%%%%%%%%%%%%%%%%%%%%%%%%%%%%%%
\section*{Appendix}
%%%%%%%%%%%%%%%%%%%%%%%%%%%%%%%

{\bf Proof of Proposition~\ref{T0}.}

To compute the integral $\int [f(\bz)]^{\alpha}d\bz$, we use the change of variables
$\bJ_{\alpha}=\alpha^{-1}\bJ$ and $\bZ_0=\bJ_{\alpha}^{-1/2}(\bZ-\bmu)$,
$\bZ_0\sim SN_d(\bzero,\bI_d,\tilde{\bfeta})$, $\tilde{\bfeta}=\bJ_{\alpha}^{1/2}\bfeta$.
We shall use the fact that $|\bJ_{\alpha}|=\alpha^{-d}|\bJ|$ for $d$-dimensional
matrices \citep{Nielsen_Nock_2012}. Then, according to Lemma~2 of \citet{Arellano-Valle_et_al_2013},
the integral $\int [f(\bz)]^{\alpha}d\bz$ should be rewritten in terms of an expected value
with respect to a standardized gaussian density as
\begin{eqnarray*}
\int [f(\bz)]^{\alpha}d\bz&=&\frac{2^{\alpha}}{|\bJ|^{\alpha/2}}|\bJ_{\alpha}|^{1/2}(2\pi)^{(1-\alpha)d/2}
\langle[\Phi_1(\tilde{\bfeta}^{\top}\bZ_{0})]^{\alpha}\rangle\\
&=&\frac{2^{\alpha}}{\alpha^{d/2}}(2\pi)^{(1-\alpha)d/2}|\bJ|^{(1-\alpha)/2}\langle[\Phi_1(W)]^{\alpha}\rangle.
\end{eqnarray*}
where $W\sim SN_1(\bzero,\|\tilde{\bfeta}\|^2,\|\tilde{\bfeta}\|)$ with $\|\tilde{\bfeta}\|=\tilde{\bfeta}^{\top}\tilde{\bfeta}$
\citep{Contreras-Reyes_Arellano-Valle_2012,Arellano-Valle_et_al_2013}, i.e., the expected value
$\langle[\Phi_1(\tilde{\bfeta}^{\top}\bZ_{0})]^{\alpha}\rangle$ is reduced from $d$ dimensions to one dimension
\citep{Arellano-Valle_et_al_2013,Contreras-Reyes_2014}. By Lemma~\ref{Pf} and setting $\bmu=\bzero$,
$\bJ=\|\tilde{\bfeta}\|^2$, $\bD=\|\tilde{\bfeta}\|$, $r=\alpha$, $\bA=s=h(w)=1$; we obtain $\tilde{\bA}=\bI_{\alpha+1}$
and $\tilde{\bD}=({\bf 1}_{\alpha},\|\tilde{\bfeta}\|)^{\top}$. Therefore, the expected value of the integral is reduced to
$$\langle[\Phi_1(W)]^{\alpha}\rangle=\frac{\Phi_{\alpha+1}(\bzero;\bzero,\bI_{\alpha+1}+
\|\tilde{\bfeta}\|^2\tilde{\bD}^{\top}\tilde{\bD})}{\Phi_1(0;0,1+\|\tilde{\bfeta}\|^4)}.\,\,\,\Box$$

{\bf Proof of Corollary~\ref{NRH}}

\begin{itemize}
\item[(i)] Follows from (\ref{rn}) and Proposition~\ref{T0}.
\item[(ii)] See Proposition 2 of \citet{Arellano-Valle_et_al_2013}.
\item[(iii)] Right side: see \citet{Contreras-Reyes_Arellano-Valle_2012}. Left side:
consider the nonsymmetrical entropy of \citet{Liu_2009} given by
$$S(\bu)=-\int f(\bu)\ln[\beta(\bu)f(\bu)]\,d\bu,$$
where $f(\bu)$ is the probability density function
of a gaussian variable $\bu$. By choosing $\beta(\bu)=2\Phi_1(\bfeta^{\top}
\bJ^{-1/2}(\bu-\bmu))$, $\bu=\bZ_N$, it follows that
$\langle{\rm\ln}\beta(\bZ_N)\rangle={\rm\ln}[2]+\Phi_1(0)=(1/2)\,{\rm\ln}(4e)$
\citep[see Proposition~4 of ][]{Azzalini_Dalla-Valle_1996}. Then, as
$\langle{\rm\ln}\beta(\bZ)\rangle\leq 2\langle{\rm\ln}\beta(\bZ_N)\rangle$,
the result is obtained.
\item[(iv)] Follows from properties (i), (ii) and (\ref{red}). $\Box$\\
\end{itemize}

{\bf Proof of Proposition~\ref{T2}}

By (\ref{dmesn}), $\phi_d(\by;\bmu,\bJ)=|\bJ|^{-1/2}\phi_d\left(\bJ^{-1/2}(\by-\bmu)\right)$, where $\phi_d(\bz)$ is
the probability density function of $N_d({\bf 0},\bI_d)$. Then, as in (\ref{T0}),
to compute the integral $\int [f(\bz)]^{\alpha}d\bz$ we use the change of variables
$\bJ_{\alpha}=\alpha^{-1}\bJ$ and $\bZ_0=\bJ_{\alpha}^{-1/2}(\bZ-\bmu)$.
In this case, $\bZ_0\sim ESN_d(\bzero,\bI_d,\tilde{\bfeta},\tau)$ with $\tilde{\bfeta}=
\bJ_{\alpha}^{1/2}\bfeta$. We shall use the fact that $|\bJ_{\alpha}|=\alpha^{-d}
|\bJ|$ for $d$-dimensional matrices \citep{Nielsen_Nock_2012}. Then, according to Lemma~2 of
\citet{Arellano-Valle_et_al_2013}, the integral $\int [f(\bz)]^{\alpha}d\bz$ should be
rewritten in terms of an expected value with respect to a standardized gaussian density as
\begin{eqnarray*}
\int [f(\bz)]^{\alpha}d\bz&=&\frac{1}{[\Phi_1(\tau)]^{\alpha}}|\bJ|^{-\frac{\alpha}{2}}|\bJ_{\alpha}|^{1/2}
(2\pi)^{(1-\alpha)\frac{d}{2}}\langle[\Phi_1(\tilde{\bfeta}^{\top}\bz_{0}+\tilde{\tau})]^{\alpha}\rangle\\
&=&\frac{1}{[\Phi_1(\tau)]^{\alpha}}\alpha^{-d}(2\pi)^{(1-\alpha)d/2}|\bJ|^{(1-\alpha)/2}\langle[\Phi_1(W)]^{\alpha}\rangle.
\end{eqnarray*}
where $W=\tilde{\bfeta}^{\top}\bZ_0+\tilde{\tau}\sim ESN_1(\tilde{\tau},\|\tilde{\bfeta}\|^2,\|\tilde{\bfeta}\|,\tau)$ with
$\|\tilde{\bfeta}\|=\tilde{\bfeta}^{\top}\tilde{\bfeta}$ \citep{Contreras-Reyes_Arellano-Valle_2012,Arellano-Valle_et_al_2013},
i.e., the expected value $\langle[\Phi_1(\tilde{\bfeta}^{\top}\bz_{0}+\tilde{\tau})]^{\alpha}\rangle$ is reduced from $d$
dimensions to one dimension \citep{Arellano-Valle_et_al_2013,Contreras-Reyes_2014}. $\Box$\\

{\bf Proof of Corollary~\ref{NHESN}}

\begin{itemize}
\item[(i)] From Proposition~\ref{T2}, we obtain directly
\begin{eqnarray*}
R_{\alpha}[f]&=&\frac{1}{1-\alpha}\left(\ln[\psi_{\alpha,d}(\bJ)]-\alpha\ln[2\Phi_1(\tau)]
+ \ln[\langle[\Phi_1(W)]^{\alpha}\rangle]\right),\\
&=&R_{\alpha}[f_0] + \frac{\alpha}{1-\alpha}\ln\left[\frac{1}{\Phi_1(\tau)}\right] +
\frac{1}{1-\alpha}\ln[\langle[\Phi_1(W)]^{\alpha}\rangle].
\end{eqnarray*}
\item[(ii)] Considering Jensen's inequality, we obtain $\langle[\Phi_1(W)]^{\alpha}\rangle\geq[\Phi_1(\langle W\rangle)]^{\alpha}$.
Then, (ii) is straightforward from (\ref{esn-mom1}).
\item[(iii)] By (\ref{red}), it follows that
$$S[f]=-\langle\ln\left[\phi_d(\bZ_0)\frac{\Phi_1(\tilde{\bfeta}^{\top}\bZ_0 + \tilde{\tau})}{\Phi_1(\tau)}\right]\rangle
=S[f_0] - \langle{\rm \ln}\left[\frac{\Phi_1(W)}{\Phi_1(\tau)}\right]\rangle,$$
where, as in Proposition~\ref{T2}, $\bZ_0=\bJ^{-1/2}(\bZ-\bmu)\sim ESN_d(\bzero,\bI_d,\tilde{\bfeta},\tau)$ and
$W={\tilde{\bfeta}}^{\top}\bZ_0+\tilde{\tau}\sim ESN_1(\tilde{\tau},\|\tilde{\bfeta}\|^2,\|\tilde{\bfeta}\|,\tau)$.
\item[(iv)] Right side: by \citet{Cover_Thomas_2006}, for any density $g(\bx)$ of a random vector $\bx\in\Delta\subset\mathbb{R}^d$
(not necessary gaussian) with zero mean and variance $\bJ=\langle\bX\bX^{\top}\rangle$, the Shannon entropy of $\bx$ is
maximized under gaussianity as $S[g] \leq (1/2)\ln[(2\pi e)^d|\bJ|]$. Then, the result is obtained
from (\ref{esn-mom2}). Left side: as in Corollary~\ref{NRH} (iii), by choosing $\beta(\bu)=\Phi_1(\bfeta^{\top}\bJ^{-1/2}
(\bu-\bmu)+\tilde{\tau})/\Phi_1(\tau)$ in the nonsymmetrical entropy, it follows that
$$\langle{\rm\ln}\beta(\bZ_N)\rangle=\Phi_1\left(\frac{\tilde{\tau}}{\sqrt{1+\|\bfeta\|}}\right)-{\rm\ln}[\Phi_1(\tau)]$$
\citep[see Proposition~4 of ][]{Azzalini_Dalla-Valle_1996}.
Then, as $\langle{\rm \ln}\beta(\bZ)\rangle\leq\langle{\rm\ln}\beta(\bZ_N)\rangle/\Phi_1(\tau)$, the result is obtained.
\item[(v)] Follows from properties (i), (iii) and (\ref{red}). $\Box$\\
\end{itemize}

{\bf Proof of Proposition~\ref{T1}}

By (\ref{TSN}), it follows that
$$\int_a^b [g(w)]^{\alpha}dw=\frac{1}{([F(z)]_a^b)^{\alpha}}\int_a^b [f(w)]^{\alpha}dw$$
and, by Proposition~\ref{T0}, the integral $\int_{a}^{b} [f(w)]^{\alpha}dw$ should be
rewritten in terms of an expected value as
$$\int_a^b [f(w)]^{\alpha}dw=\psi_{\alpha,1}(J) \langle[\Phi_1(u)]^{\alpha}|a_0<u\leq b_0\rangle,$$
where $U\sim SN_1(0,\tilde{\lambda}^2,\tilde{\lambda})$, $\tilde{\lambda}^2=\omega\lambda^2/\alpha$,
$a_0=\lambda(a-\mu)/\omega$ and $b_0=\lambda(b-\mu)/\omega$.
Again, by Lemma~\ref{Pf} and setting $\bmu=0$, $J=\tilde{\lambda}^2$, $r=\alpha$,
$d=s=\bA=h(u)=1$; we obtain $\tilde{\bA}=\bI_{\alpha+1}$, $\tilde{\bD}=({\bf 1}_{\alpha},\tilde{\lambda})^{\top}$
and $\tilde{\bJ}=\bI_{\alpha+1}+\tilde{\lambda}^2\tilde{\bD}^{\top}\tilde{\bD}$. Then, the expected value is
\begin{eqnarray*}
\langle[\Phi_1(u)]^{\alpha}|a_0<u\leq b_0\rangle&=&2\Phi_{\alpha+1}(\bzero;\bzero,\tilde{\bJ})[H(v)]_{a_0}^{b_0},
\end{eqnarray*}
where $H(v)$ is the cumulative density function of a closed skew-gaussian variable $V\sim CSN_{1,2}(0,\tilde{\lambda}^2,\tilde{\bB},\bzero,\bI_2)$
with $\tilde{\bB}=(1,\tilde{\lambda})^{\top}$ \citep[see Proposition~3 of ][]{Flecher_et_al_2010}. $\Box$\\

\section*{Acknowledgements}

This work was supported by Instituto de Fomento Pesquero (IFOP, {\tt http://www.ifop.cl/}), Valpara\'iso, Chile.
The author would like to thank the editor and an anonymous referee for their helpful comments and suggestions.\\


\begin{thebibliography}{100}

\bibitem[Azzalini(1985)]{Azzalini_1985}
Azzalini, A., 1985.
A Class of Distributions which includes the Normal Ones.
Scand. J. Stat. 12, 171-178.

\bibitem[Azzalini and Dalla-Valle(1996)]{Azzalini_Dalla-Valle_1996}
Azzalini, A., Dalla-Valle, A., 1996.
The multivariate skew-normal distribution.
Biometrika 83, 715-726.

\bibitem[Azzalini and Capitanio(1999)]{Azzalini_Capitanio_1999}
Azzalini, A., Capitanio, A., 1999.
Statistical applications of the multivariate skew normal distributions.
J. Roy. Stat. Soc. Ser. B 61, 579-602.

\bibitem[Azzalini(2013)]{Azzalini_2013}
Azzalini, A., 2013.
The Skew-Normal and Related Families.
Vol. 3, Cambridge University Press.

\bibitem[Gonz\'alez-Far\'ias et al.(2004)]{Gonzalez-Farias_et_al_2004}
Gonz\'alez-Far\'ias, G., Dom\'inguez-Molina, J., Gupta, A., 2004.
Additive properties of skew normal random vectors.
J. Stat. Plann. Inference 126, 521-534.

\bibitem[Rezaie et al.(2014)]{Rezaie_et_al_2014}
Rezaie, J., Eidsvik, J., Mukerji, T., 2014.
Value of information analysis and Bayesian inversion for closed skew-normal distributions:
Applications to seismic amplitude variation with offset data.
Geophys. 79, R151-R163.

\bibitem[Capitanio et al.(2003)]{Capitanio_et_al_2003}
Capitanio, A., Azzalini, A., Stanghellini, E., 2003.
Graphical models for skew-normal variates.
Scand. J. Stat. 30, 129-144.

\bibitem[Flecher et al.(2010)]{Flecher_et_al_2010}
Flecher, C., Allard, D., Naveau, P., 2010.
Truncated skew-normal distributions: moments, estimation by weighted moments and application to climatic data.
Metron 68, 265-279.

\bibitem[Contreras-Reyes and Arellano-Valle(2012)]{Contreras-Reyes_Arellano-Valle_2012}
Contreras-Reyes, J.E., Arellano-Valle, R.B., 2012.
Kullback-Leibler divergence measure for Multivariate Skew-Normal Distributions.
Entropy 14, 1606-1626.

\bibitem[Arellano-Valle et al.(2013)]{Arellano-Valle_et_al_2013}
Arellano-Valle, R.B., Contreras-Reyes, J.E., Genton, M.G., 2013.
Shannon entropy and mutual information for multivariate skew-elliptical distributions.
Scand. J. Stat. 40, 42-62.

\bibitem[Figiel(2014)]{Figiel_2014}
Figiel, {\L}., 2014.
Effect of the interphase on large deformation behaviour of polymer--clay nanocomposites near the glass transition: 2D RVE computational modelling.
Comput. Mater. Sci. 84, 244-254.

\bibitem[Zhou and Wang(2008)]{Zhou_Wang_2008}
Zhou, J., Wang, X., 2008.
Accurate closed-form approximation for pricing Asian and basket options.
Appl. Stochastic Models Bus. Ind. 24, 343-358.

%\bibitem[Flecher et al.(2010b)]{Flecher_et_al_2010b}
%Flecher, C., Naveau, P., Allard, D., Brisson, N., 2010b.
%A stochastic daily weather generator for skewed data.
%Water Resour. Res. 46.

\bibitem[Genton(2004)]{Genton_2004}
Genton, M.G., 2004.
Skew-elliptical distributions and their applications: A journey beyond normality.
Chapman \& Hall/CRC, Boca Raton, FL.

\bibitem[Contreras-Reyes(2014)]{Contreras-Reyes_2014}
Contreras-Reyes, J.E., 2014.
Asymptotic form of the Kullback-Leibler divergence for multivariate asymmetric heavy-tailed distributions.
Physica A 395, 200-208.

\bibitem[R\'enyi(1970)]{Renyi_1970}
R\'enyi, A., 1970.
Probability theory.
North-Holland, Amsterdam.

\bibitem[L\'opez-Ruiz et al.(1995)]{Lopez-Ruiz_et_al_1995}
L\'opez-Ruiz, R., Mancini, H.L., Calbet, X., 1995.
A statistical measure of complexity.
Phys. Lett. A 209, 321-326.

\bibitem[Anteneodo and Plastino(1996)]{Anteneodo_Plastino_1996}
Anteneodo, C., Plastino, A.R., 1996.
Some features of the L\'opez-Ruiz-Mancini-Calbet (LMC) statistical measure of complexity.
Phys. Lett. A 223, 348-354.

\bibitem[Yamano(2004)]{Yamano_2004}
Yamano, T., 2004.
A statistical measure of complexity with nonextensive entropy.
Physica A 340, 131-137.

\bibitem[Carpi et al.(2011)]{Carpi_et_al_2011}
Carpi, L.C., Rosso, O.A., Saco, P.M., Ravetti, M.G., 2011.
Analyzing complex networks evolution through Information Theory quantifiers.
Phys. Lett. A 375, 801-804.

\bibitem[S\'anchez-Moreno et al.(2014)]{Sanchez-Moreno_et_al_2014}
S\'anchez-Moreno, P., Angulo, J.C., Dehesa, J.S., 2014.
A generalized complexity measure based on R\'enyi entropy.
Eur. Phys. J. D 68, 212.

\bibitem[Golshani and Pasha(2010)]{Golshani_Pasha_2010}
Golshani, L., Pasha, E., 2010.
R\'enyi entropy rate for Gaussian processes.
Inform. Sci. 180, 1486-1491.

\bibitem[Liu et al.(2012)]{Liu_et_al_2012}
Liu, T., Zhang, P., Dai, W-.S., Xie, M., 2012.
An intermediate distribution between Gaussian and Cauchy distributions.
Physica A 391, 5411-5421.

\bibitem[Cover and Thomas(2006)]{Cover_Thomas_2006}
Cover, T.M., Thomas, J.A., 2006.
Elements of information theory.
Wiley \& Son, Inc., New York, NY, USA.

\bibitem[Dembo et al.(1991)]{Dembo_et_al_1991}
Dembo, A., Cover, T.M., Thomas, J.A., 1991.
Information Theoretic Inequalities.
IEEE Trans. Inform. Theory 37, 1501-1518.

\bibitem[Flecher et al.(2009)]{Flecher_et_al_2009}
Flecher, C., Naveau, P., Allard, D., 2009.
Estimating the Closed Skew-Normal distributions parameters using weighted moments.
Stat. Prob. Lett. 79, 1977-1984.

\bibitem[Pourahmadi(2007)]{Pourahmadi_2007}
Pourahmadi, M., 2007.
Skew-Normal ARMA Models with Nonlinear Heteroscedastic Predictors.
Commun. Stat. A-Theor. 36, 1803-1819.

\bibitem[Owen(1956)]{Owen_1956}
Owen, D.B., 1956.
Tables for computing bivariate normal probabilities.
Ann. Math. Stat. 27, 1075-1090.

\bibitem[Nielsen and Nock(2012)]{Nielsen_Nock_2012}
Nielsen, F., Nock, R., 2012.
A closed-form expression for the Sharma--Mittal entropy of exponential families.
J. Phys. A: Math. Theor. 45, 032003.

\bibitem[Liu(2009)]{Liu_2009}
Liu, C.-S., 2009.
Nonsymmetric entropy and maximum nonsymmetric entropy principle.
Chaos Soliton. Fract. 40, 2469-2474.

\end{thebibliography}
\end{document}